\documentclass[twocolumn,showpacs,preprintnumbers,amsmath,amssymb]{revtex4}

\usepackage{graphicx}
\usepackage{dcolumn}
\usepackage{bm}
\usepackage[colorlinks=true, citecolor=red, linkcolor=blue, urlcolor=blue ]{hyperref}

\newcommand\etal{\mbox{\textit{et al.}}}

\renewcommand{\vec}[1]{\mbox{\boldmath$ #1 $}}

\usepackage{lineno}
\usepackage[usenames]{color}

\begin{document}


\title{Instability of a jet of active fluid composed of microswimmers}

\title{Instability of  an active fluid jet}

\author{Takuji Ishikawa$^{1,*}$}
\author{Thanh-Nghi Dang$^{2}$}
\author{Eric Lauga$^{3}$}
\affiliation{$^1$Department of Biomedical Engineering, Tohoku University, Sendai 980-8579, Japan\\
$^2$Faculty of Mechanical Engineering, RWTH Aachen University, 52056 Aachen, Germany\\
$^3$Department of Applied Mathematics and Theoretical Physics, University of Cambridge, Cambridge CB3 0WA, UK\\
}%

\date{\today}

\begin{abstract}

The breakup of a fluid jet into droplets has long fascinated natural scientists, with early research dating back to the 19th century. Infinitesimal perturbations to a jet grow because of   surface tension, which eventually leads to breakup of the jet into droplets (Rayleigh-Plateau instability). Although this classical phenomenon has long been studied, it is not clear how it is modified when the fluid is replaced by an \emph{active fluid}. In this study, we investigate instabilities of a jet of an active fluid. The active fluid is modelled by a suspension of microswimmers that propel themselves by generating surface tangential velocities, i.e.~squirmers. The squirmers are assumed to be bottom-heavy and heavier than the surrounding fluid, so that a downward jet of the active fluid self-assembles under gravity. Hydrodynamic interactions between squirmers are computed using the Stokesian dynamics method,
in which near-field hydrodynamics are accurately calculated.
We find that jets of active fluids are unstable, with
different
unstable modes  between pullers and pushers. For an active fluid of pullers, the jet breaks up into droplets in a varicose manner reminiscent of a Newtonian fluid. For   pushers, on the other hand, the jet buckles and undergoes a waving (sinuous) instability. The physical mechanisms for these two instabilities can be understood by an inspection of the stress fields in the jets and parametric study reveals the importance of hydrodynamic interactions in the instabilities.
Although both gravity and bottom-heaviness play an essential role in realizing the downward jet, their influence on the instability   is found to be limited.
Our findings help reveal new features of  the collective properties of active fluids.
\end{abstract}

\maketitle


\section{Introduction}

The breakup of a fluid jet into droplets -- so-called Rayleigh-Plateau instability -- has long fascinated scientists, with early research dating back to the 19th century~\cite{Savart}. Plateau put forward the energetic arguments allowing to understand the instability as a release of surface energy while 
Rayleigh explained  the instability   jet   as the balance between surface tension  and inertia in the fluid. Rayleigh derived that the optimal wavelength, at which perturbations grow fastest, is about ninefold   the jet radius, which sets the typical size of droplets~\cite{Rayleigh}. The time scale $T$ on which perturbations grow and eventually break the jet can also be derived as $T \sim \sqrt{r^3 \rho / \gamma}$, where $r$ is the jet radius, $\rho$ is the density and $\gamma$ is the surface tension~\cite{Eggers}. A large number of experimental, theoretical and computational studies have been performed since then, and the breakup of a Newtonian jet is now a classical problem~\cite{Lasheras,Eggers2,Polanco}. Although the phenomenon has been well studied, it is not clear how it is modified when the fluid is replaced by an \emph{active fluid}. In this study, therefore, we investigate instabilities of a jet of an active fluid.

An active fluid denotes a collection of particles that propel themselves by consuming free energy available in their environment~\cite{Marchetti}. Examples include suspensions of microorganisms, the cytoskeleton of eukaryotic cells and aquatic flocks. Active fluids have attracted physicists, biologists and engineers~\cite{Ramaswamy,Saintillan}, and deciphering their complex collective dynamics may lead to a better understanding of biological systems and an application in bioengineering~\cite{Doostmohammadi,Klotsa}. Past studies have shown that the properties of active fluids differ significantly from those of classical passive fluids. For example, active fluids show motility-induced phase separation where an assembly of active particles phase separates into dense and dilute regions~\cite{Tjhung}. Turbulent (but inertialess) motions of active fluids exhibit a different energy spectrum than that of Newtonian fluids at high Reynolds numbers~\cite{Bratanov}. The rheological properties of active fluids are also altered by the internal stress induced by energy-injecting active particles~\cite{Saintillan,Ishikawa}. An active fluid can even display a superfluid like transition, where the viscous resistance to shear vanishes~\cite{Lopez}. Recently, Jibuti \etal~\cite{Jibuti} performed simulations on a suspension of phototactic microswimmers in a Poiseuille flow. They observed an instability of the focused jet, which leads to its fractionation in clusters. These studies have revealed the complex dynamics of active fluids and motivate the further analysis of simple collective modes.

In this study, we consider as an  active fluid a suspension of microswimmers each of which self-propel by generating surface tangential velocities, i.e.~squirmers~\cite{Blake,Pedley}. The classical squirmer model has been used to investigate a variety of active suspension properties, such as rheology~\cite{Ishikawa2}, coherent structures~\cite{Ishikawa3} and mass transport~\cite{Ishikawa4}. The squirmers in this study are assumed to be bottom-heavy and heavier than the surrounding fluid, so that a downward jet of the active fluid self-assembles under gravity.
We note that both gravity and bottom-heaviness play an essential role in realizing the downward jet, but their influence on the instability  is limited, as we will discuss in sections III and IV.
The setup of downward jets
is relevant to  range of biological situations, including a falling plume of a cell suspension, i.e.~bioconvection~\cite{Pedley1992,Hill2005,Bees2020}. The basic mechanism of bioconvection is analogous to that of Rayleigh-B\'{e}ard convection, in which density instability develops when the upper cell layer become denser than the lower fluid regions. The upward swimming of cells can be induced by various mechanisms. For example, some microorganisms display gravitaxis and   
passively reorient
gravitationally upward due to their bottom-heaviness~\cite{Pedley1992} or a shape asymmetry~\cite{Kage2020}. Many microalgae can swim towards light sources for photosynthesis, called as phototaxis~\cite{Jekely,Drescher,Maleprade}. Chemotaxis is the ability for cells to swim towards chemoattractant, such as oxygen at a water-air interface~\cite{Kaupp,Stocker}. The focusing of cells can also be generated by introducing an external flow field. In a landmark paper, Kessler~\cite{Kessler} showed that upswimming microalgae in pipe flow focus to the pipe axis, a phenomenon now referred to as gyrotaxis.

Previously, instabilities arising in  a single line of aligned swimmers moving along the same direction were analyzed. When the swimmers are pullers, i.e.~when their thrust is generated in front of the body, a clustering instability was observed~\cite{Lauga1}. In contrast, when the swimmers are pushers, i.e.~with a thrust generated behind the body, a zigzag instability took place~\cite{Lauga2}. These results revealed that the instabilities  depend strongly on the type of swimmers. A natural extension of these previous studies is to investigate dense lines of swimmers, i.e.~jets, instead of single lines. Therefore, this paper aims to provide a comprehensive picture of instabilities of an active fluid jet by performing parametric simulations, including both pullers and pushers.
The essential difference between these former theoretical studies and the present work is that our approach includes both near-field hydrodynamic interactions and excluded volume effects. In Refs.~\cite{Lauga1, Lauga2} the microswimmers were modelled as a point stresslets, whereas in Ref.~\cite{Jibuti} the microswimmer was modelled as a set of three point forces. Therefore, these former studies were valid only in the far field (i.e.~in the dilute limit) in the absence of near-field hydrodynamic interactions and  excluded volume effects. In the present work, on the other hand, we use the Stokesian Dynamics method, in which near-field lubrication forces and  excluded volume effects are precisely calculated. Hence, we are able to deal with non-dilute suspensions of squirmers and tackle the stability of clustered suspensions.

The structures of our paper is as follows. In \S II, we summarise the squirmer model, basic equations and numerical methods. Hydrodynamic interactions between squirmers are computed using the Stokesian dynamics method. The clustering instability induced by pullers is investigated in \S III while the waving instability induced by pushers is address in  \S IV. Our   results are finally discussed and summarised in \S V.


\section{Squirmer model and numerical method}

\subsection{Squirmer model}

The active fluid is modelled by a suspension of spherical squirmers~\cite{Blake,Pedley}. The squirmer is assumed to be heavier than the surrounding fluid, bottom-heavy and non-Brownian, and to swim at a very small Reynolds number. The sphere's surface is assumed to force the surrounding fluid purely tangentially, and these tangential motions are assumed to be axisymmetric and time-independent. We follow Ref.~\cite{Ishikawa5}, and the tangential surface velocity on a squirmer is prescribed to be $u_s = (3U_0/2)(\sin \theta + \beta \sin \theta \cos \theta)$, where $U_0$ is the swimming speed of an isolated squirmer, and $\theta$ is the angle from the orientation vector. The first terms gives rise to swimming while the second term controls the far-field flow in the shape of a force dipole, or stresslet.  The parameter $\beta$ can have either sign; a squirmer with  $\beta > 0$ is a puller, whereas a squirmer   $\beta < 0$ is a pusher.

\begin{figure}
\begin{center}
\centerline{\includegraphics[scale=0.45]{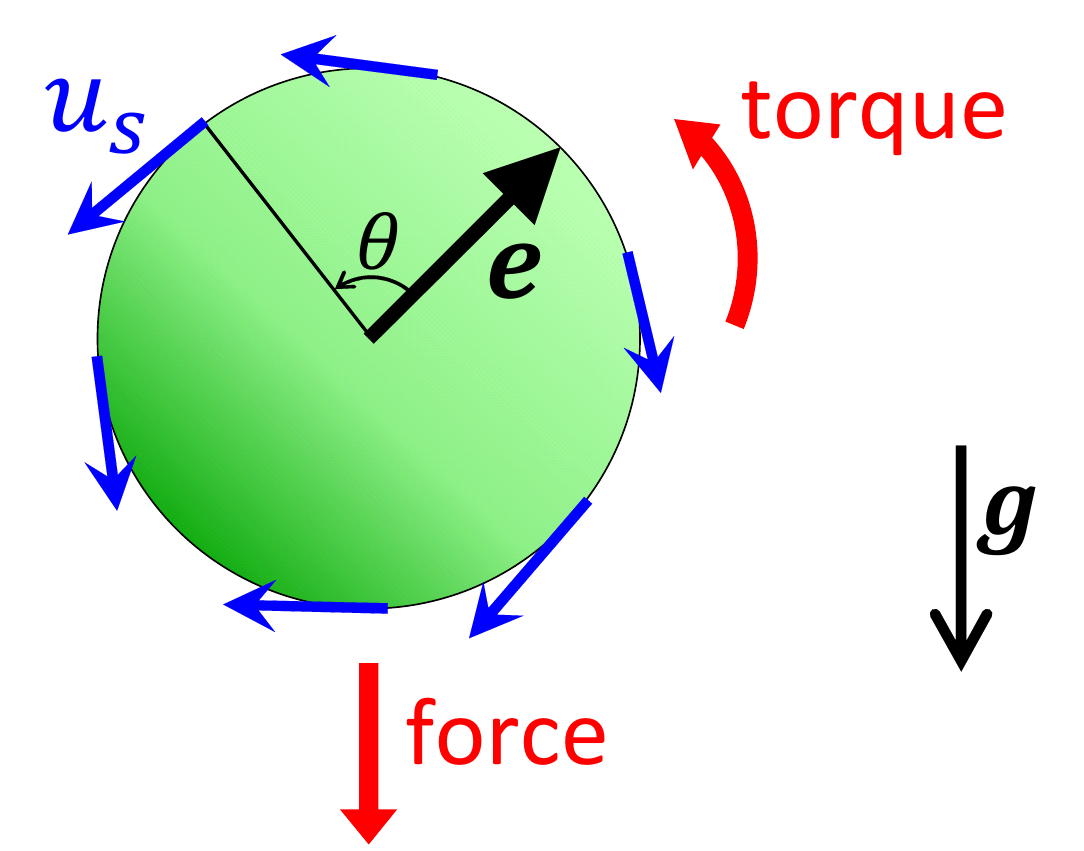}}
\caption{Squirmer model with orientation vector $\vec{e}$ and surface velocity $u_s$. External force and torque are exerted under the gravity field $\vec{g}$.}
\label{fig0}
\end{center}
\end{figure}

Squirmers are denser than the surrounding fluid, and an external force of $\frac{4}{3}\pi a^3 \Delta \rho \vec{g}$ is exerted on each particle, where $a$ is the radius, $\Delta \rho$ is the density mismatch. Squirmers do tend to align in the direction opposite to gravity due to their bottom-heaviness. The restoring torque is given as $\frac{4}{3}\pi a^3 \rho h \vec{e} \times \vec{g}$, where $\rho$ is the density, $h$ is the distance between  gravity and geometric centers, $\vec{e}$ is the orientation vector, $\vec{g}$ is the gravitational acceleration vector (see Fig.~\ref{fig0}).

\begin{figure*}
\begin{center}
\centerline{\includegraphics[scale=0.48]{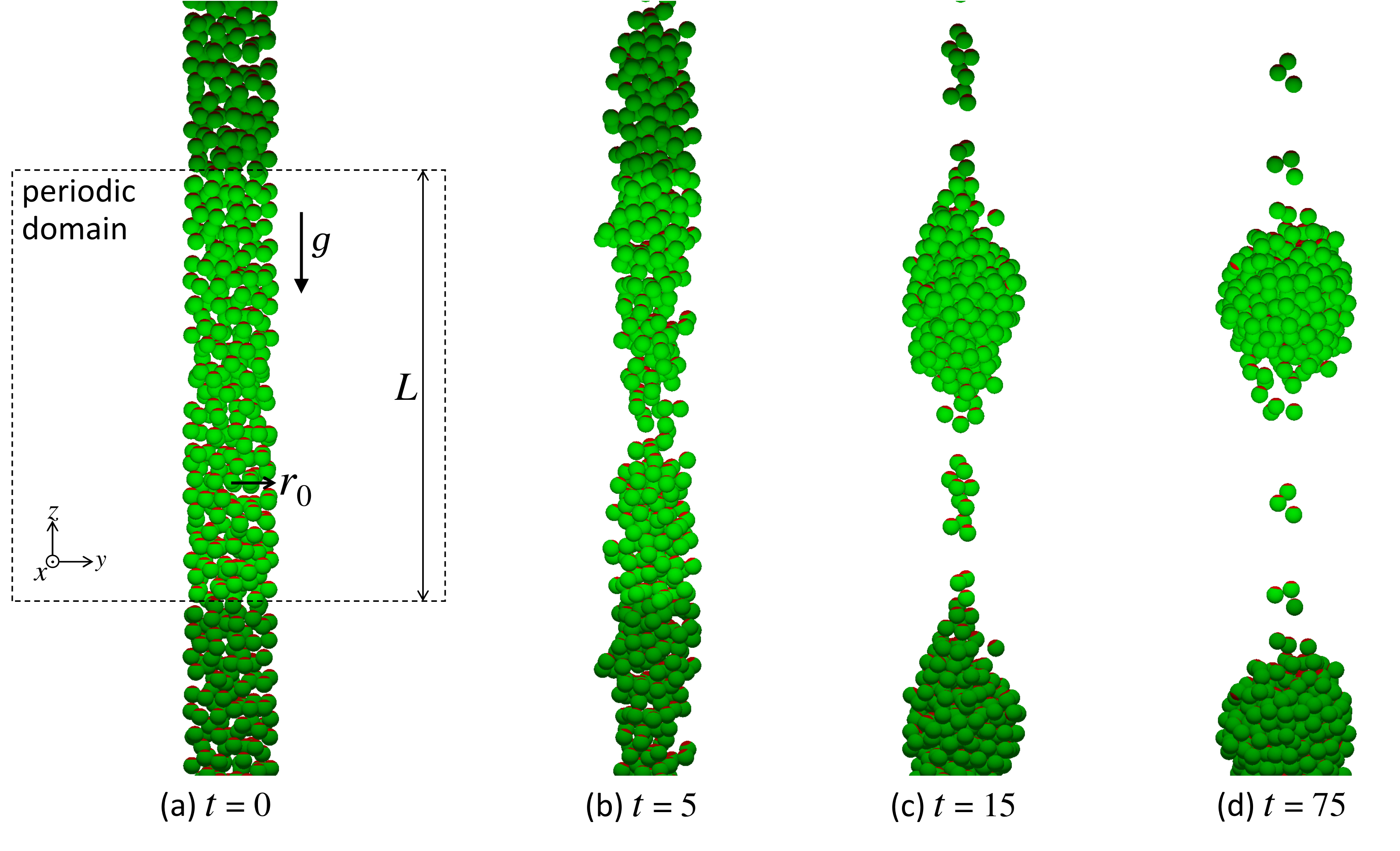}}
\caption{Clustering instability observed for pullers ($\beta = 1, F_g = 5\pi, G_{bh} = 100$ and $N=300$), where $t$ is non-dimensionalised by $a/U_0$; see also supplementary Movie 1~\cite{SM}. (a) Initially, 300 squirmers are placed in a column of radius $r_0$ and oriented vertically upward, where   gravity acts in the $-z$ direction. The anterior part of each squirmer is colored in red. The periodic domain is a cube with side length $L \approx 50$, where $L$ is non-dimensionalised by $a$. (b) Jet configuration at $t = 5$. (c) Jet configuration at $t = 15$. (d) Jet configuration at $t = 75$, where a droplet is steadily formed.}
\label{fig1}
\end{center}
\end{figure*}

\subsection{Stokesian Dynamics}

The Stokesian dynamics method for computing hydrodynamic interactions among an infinite suspension of squirmers was developed by Ishikawa \etal~\cite{Ishikawa3}. The method includes an infinite number of reflected far-field interactions among an infinite number of squirmers as well as near-field lubrication forces.
Hence, this simulation method can accurately account for  near-field hydrodynamic interactions and excluded volume effects and can be used to study concentrated suspensions and clusters of squirmers; this is the main advantage of the present method over former approaches to similar problems.

The original version of the Stokesian dynamics method can be slightly simplified to deal with a larger number of particles~\cite{Ishikawa6}, following the method developed for passive particles~\cite{Banchio}. We refer the readers to Ref.~\cite{Ishikawa6} for details and we give only a rapid overview. At a negligible particle Reynolds number, the motions of $N$ squirmers periodically replicated in three-dimensional fluid otherwise at rest can be given as~\cite{Ishikawa6}
\begin{eqnarray}
\label{mat.5}
\left[ \vec{I} + \vec{M}_{FU}^{far} : \vec{K}_{FU}^{2b} \right] .
\left( \begin{array}{c}
\! \vec{U} - U_0 \vec{e} + \vec{H}_{sq} \! \\
\vec{\Omega}
\end{array} \right)
= ~~~~~~~~~~~
\nonumber \\
\nonumber
\vec{M}_{FU}^{far} .
\left\{ \!
\left( \begin{array}{c}
\! \vec{F} + \vec{F}^\alpha \! \\
\! \vec{T} + \vec{T}^\alpha \!
\end{array} \right)
\! - \vec{K}_{SU}^{2b} \! : \!
\left[ - \frac{3}{10}U_0 \beta \left( 3 \vec{e} \vec{e} - \vec{I} \right)
\right] \!
\right\} \\
+ \vec{M}_{SU}^{far} : \vec{S}^{far}~,~~~~~~~~~~~~~~~~~~~~~~~~~~~~~~~~~~~~~~~~~~~~~~
\end{eqnarray}
with
\begin{equation}
\left( \begin{array}{c}
\! \vec{F}^\alpha \! \\
\! \vec{T}^\alpha \!
\end{array} \right)
=
\vec{K}_{2B}^{near} \! : \!
\left( \begin{array}{c}
- U_0 \vec{e} + \vec{H}_{sq} \\
0 \\
\! - \frac{3}{10}U_0 \beta \left( 3 \vec{e} \vec{e} - \vec{I} \right) \! \!
\end{array} \right)
-
\left( \begin{array}{c}
\! \vec{F}_{sq}^{near} \! \! \\
\! \vec{T}_{sq}^{near} \! \!
\end{array} \right) ,
\end{equation}
where $\vec{F}$, $\vec{T}$ and $\vec{S}$ are, respectively, the force, torque, and stresslet a squirmer exerts on the fluid while $\vec{U}$ and $\vec{\Omega}$ are the translational and rotational velocities of the squirmer; $\vec{S}^{far}$ and $\vec{H}_{sq}$ are respectively the far-field contributions to the stresslet and the irreducible quadrupole; $\vec{M}^{far}$ is the far-field contribution to the grand mobility matrix derived from the Fax\'en's laws. The infinite extent of the suspension is taken into account using Ewald summation~\cite{Beenakker}.
To include near-field interactions, we add near-field multipoles in a pairwise additive fashion using the boundary element method~\cite{Ishikawa5}. A short-range inter-particle repulsive force~\cite{Ishikawa3} is added to the system to avoid the prohibitively small time step needed to overcome the problem of overlapping particles. The accuracy of the method was confirmed in Ref.~\cite{Ishikawa6}.

\subsection{Setup and numerical procedure}

Hereafter, all equations are non-dimensionalised using the radius $a$, the isolated swimming speed $U_0$ and the fluid viscosity $\mu$. As a result, a dimensionless time $t=1$ is equivalent to the dimensional time $a/U_0$. The sedimentation force is non-dimensionalised by $\mu a U_0$, so $F_g = 6 \pi$ indicates that the magnitude of sedimentation velocity is equivalent to $U_0$. The effect of bottom-heaviness is discussed by a dimensionless number $G_{bh}$, defined as $G_{bh}= 4 \pi \rho g a h/ (3 \mu U_0)$.

Initially, a jet of the active fluid is placed in the middle of a unit cubic domain with side length $L$. The jet axis is aligned with the gravity axis (i.e.~the $z$-axis). A suspension of infinite extent is represented by the triply periodic boundary conditions. The jet is thus infinitely long in the $z$-direction and repeated in the $x$- and $y$-directions with spacing $L$. Initially, the centers of the squirmers are randomly placed in a column of radius $r_0$ along the $z$-axis, with avoiding overlap of particles, as shown in Fig.~\ref{fig1}(a).
The initial volume fraction of squirmers in the cylinder column is about 20\%.
The  squirmers are all initially directed vertically upward. The dynamic motions of the squirmers  are then  calculated using the Stokesian dynamics method with a 4$^{th}$ order Adams-Bashforth time-marching scheme.

The range of parameters used in this study is as follows: the number of squirmers in the unit domain $250 \le N \le 500$; swimming mode $-4 \le \beta \le 4$; sedimentation force $0 \le F_g \le 5\pi$; bottom-heavy effect $0 \le G_{bh} \le 100$; system size $42 \le L \le 83$; and initial jet radius $3.9 \le r_0 \le 7.7$. This broad range of values is employed in order to cover a wide variety of swimmer types and jet conditions.


\begin{figure}[t]
\begin{center}
\centerline{\includegraphics[scale=0.35]{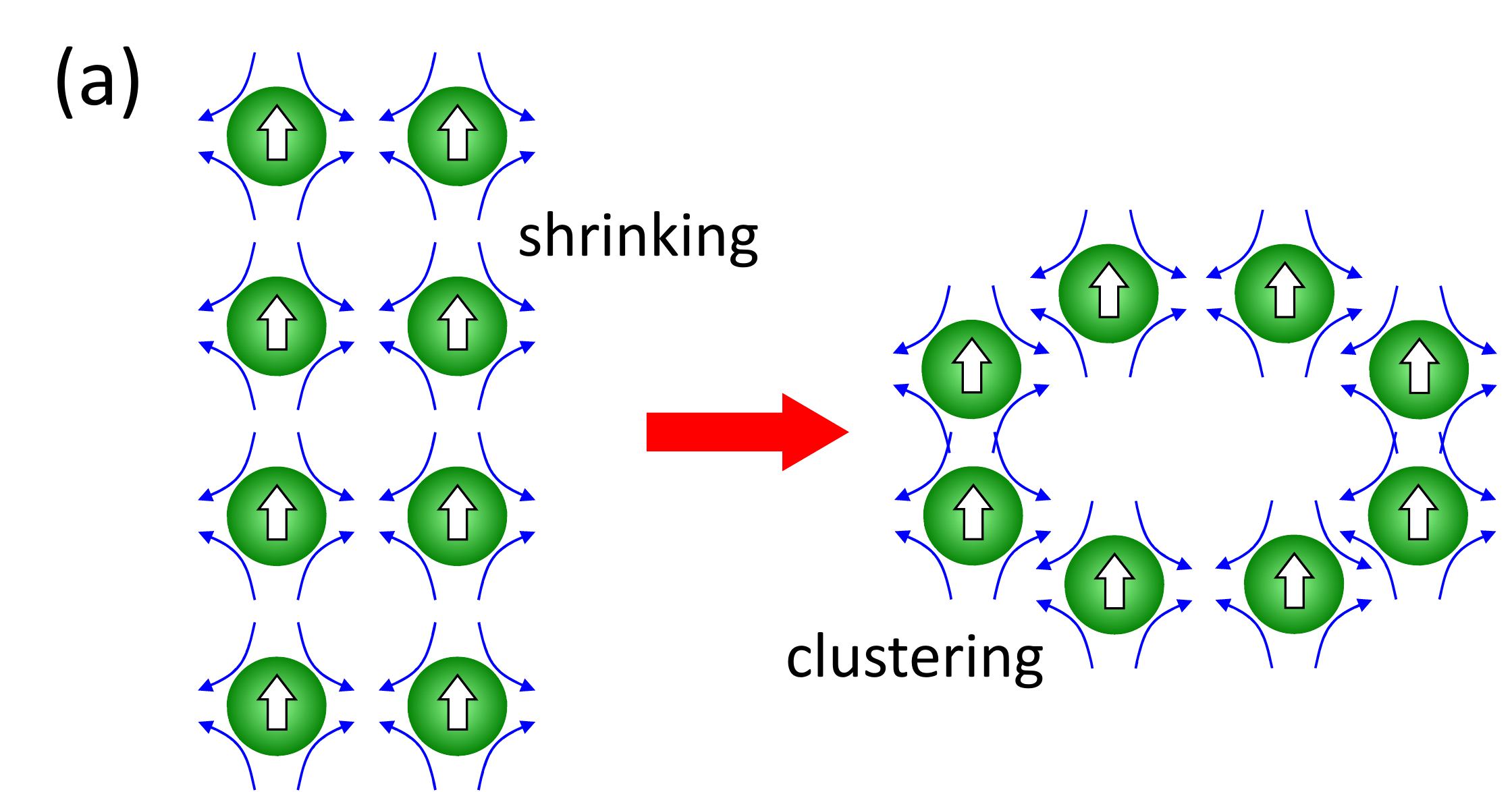}}
\vspace{10mm}
\centerline{\includegraphics[scale=0.33]{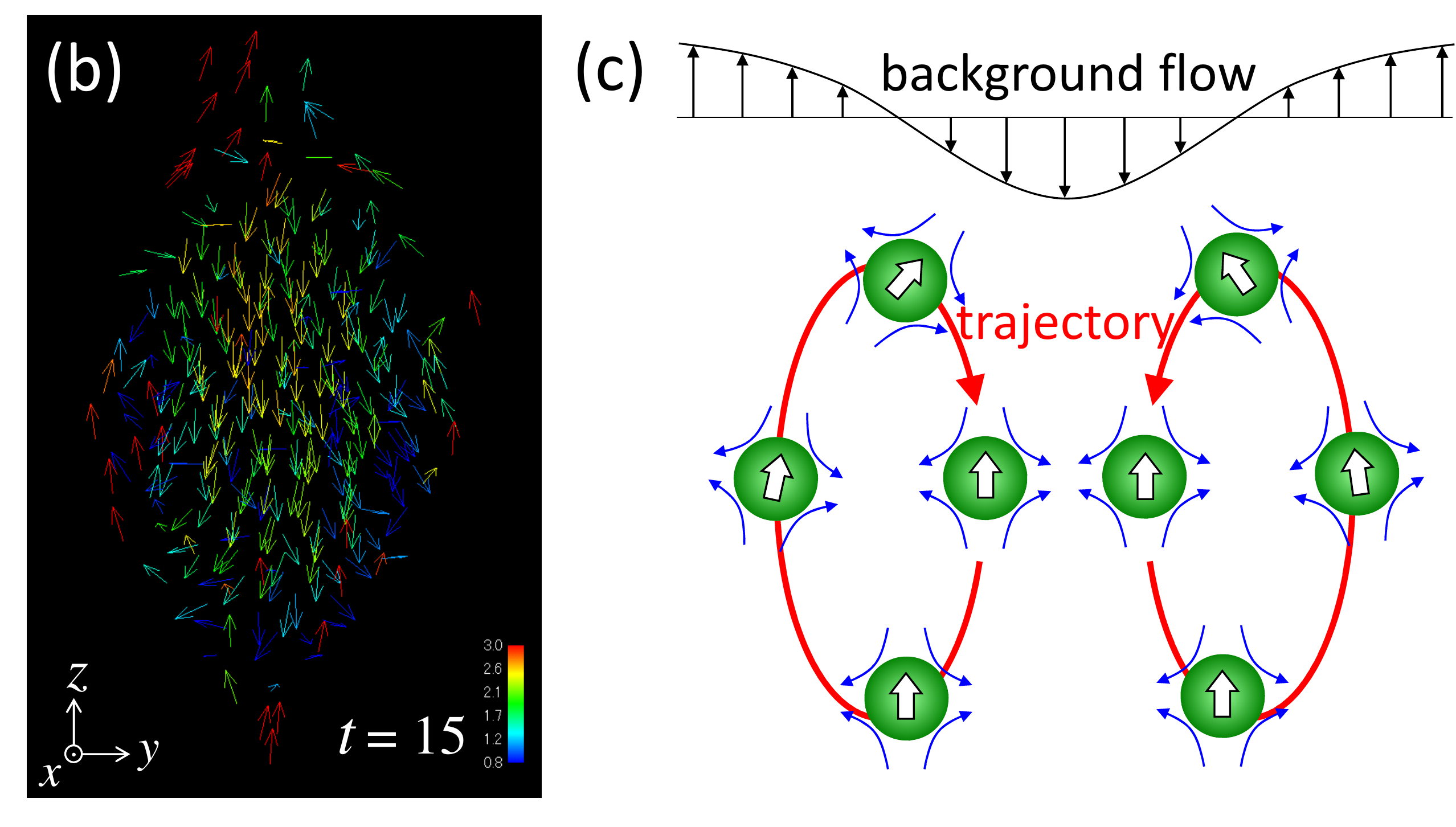}}
\caption{Formation of active droplets of pullers. (a) Schematics of the clustering instability of pullers. Squirmers vertically attract each other, and the jet shrinks to form a droplet. The white arrows indicate the orientation of the squirmers while the blue arrows show the flow generated by the pullers. (b) Velocity vectors of squirmers relative to the squirmers' average velocity  ($t = 15, \beta = 1, F_g = 5\pi, G_{bh} = 100$ and $N=300$). (c) Schematics of squirmers' motions in the background flow. Red arrows indicate the trajectories.}
\label{fig2}
\end{center}
\end{figure}

\section{Clustering instability induced by pullers}

We first investigate the clustering instability displayed by pullers. Figure \ref{fig1} shows the time change of jet configurations with the parameters $\beta = 1, F_g = 5\pi, G_{bh} = 100$ and $N=300$ (see also supplementary Movie 1~\cite{SM}). The jet sinks due to the sedimentation force, while the surrounding fluid moves upward so that there is no net flow across the horizontal plane. The background velocity field and the hydrodynamic interactions between the squirmers induce a progressive change in the  configuration  of the jet. When $t = 5$,   a narrow part appears in the middle of the jet. This varicose-like perturbation to the jet configuration grows gradually with time, and at $t = 15$ the initially continuous jet splits into droplets. These dynamical droplets are stable and can be found even at $t = 75$.  This observed phenomenon is similar in appearance to the classical (varicose) Rayleigh-Plateau instability, but of course the driving force is different (surface tension vs.~active stresses). Instead, our results are 
reminiscent of that observed for lines of puller swimmers~\cite{Lauga1}.

\begin{figure}[t]
\begin{center}
\centerline{\includegraphics[scale=0.5]{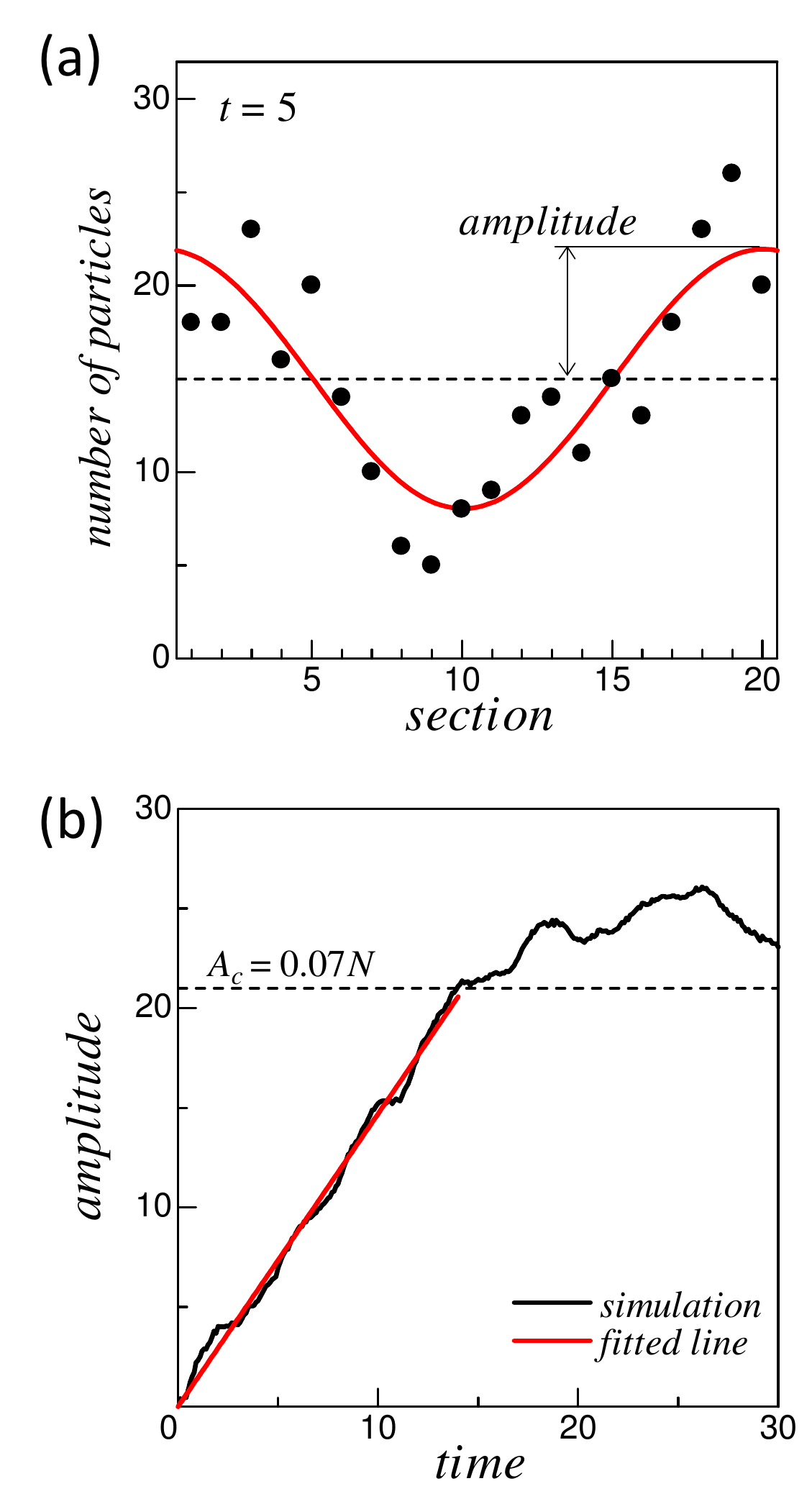}}
\caption{Growth of the clustering instability of pullers ($\beta = 1, F_g = 5\pi, G_{bh} = 100$ and $N=300$). (a) Vertical distribution of number density of particles. The plume is split into 20 sections in the vertical direction, and the number of particles in each section is plotted. The red sinusoidal curve is fitted using least squares, from which the perturbation amplitude is calculated. (b) Time evolution  of the amplitude. Threshold value $A_c$ is set as $A_c = 0.07N$, and the perturbation growth rate $V_{drop}$ is obtained by fitting the data less than $A_c$ using least squares.}
\label{fig3}
\end{center}
\end{figure}

\begin{figure}[t]
\begin{center}
\centerline{\includegraphics[scale=0.5]{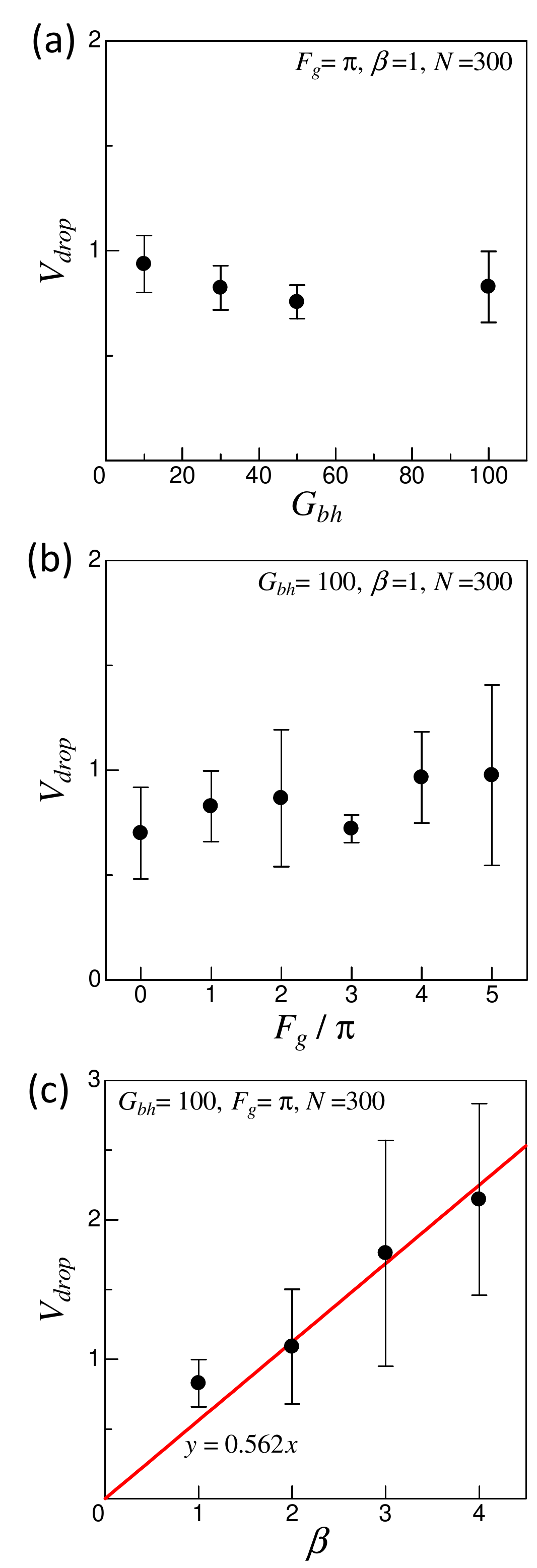}}
\caption{Effect of parameters on the growth rate $V_{drop}$ of the clustering instability for pullers ($N=300$). Error bars indicate standard deviation of five different computations. (a) Effect of $G_{bh}$ ($\beta = 1, F_g = \pi$). (b) Effect of $F_g$ ($\beta = 1, G_{bh} = 100$). (c) Effect of $\beta$ ($F_g = \pi, G_{bh} = 100$). A red line with slope 0.562 is plotted using least-squares.}
\label{fig4}
\end{center}
\end{figure}

\begin{figure}[t]
\begin{center}
\centerline{\includegraphics[scale=0.5]{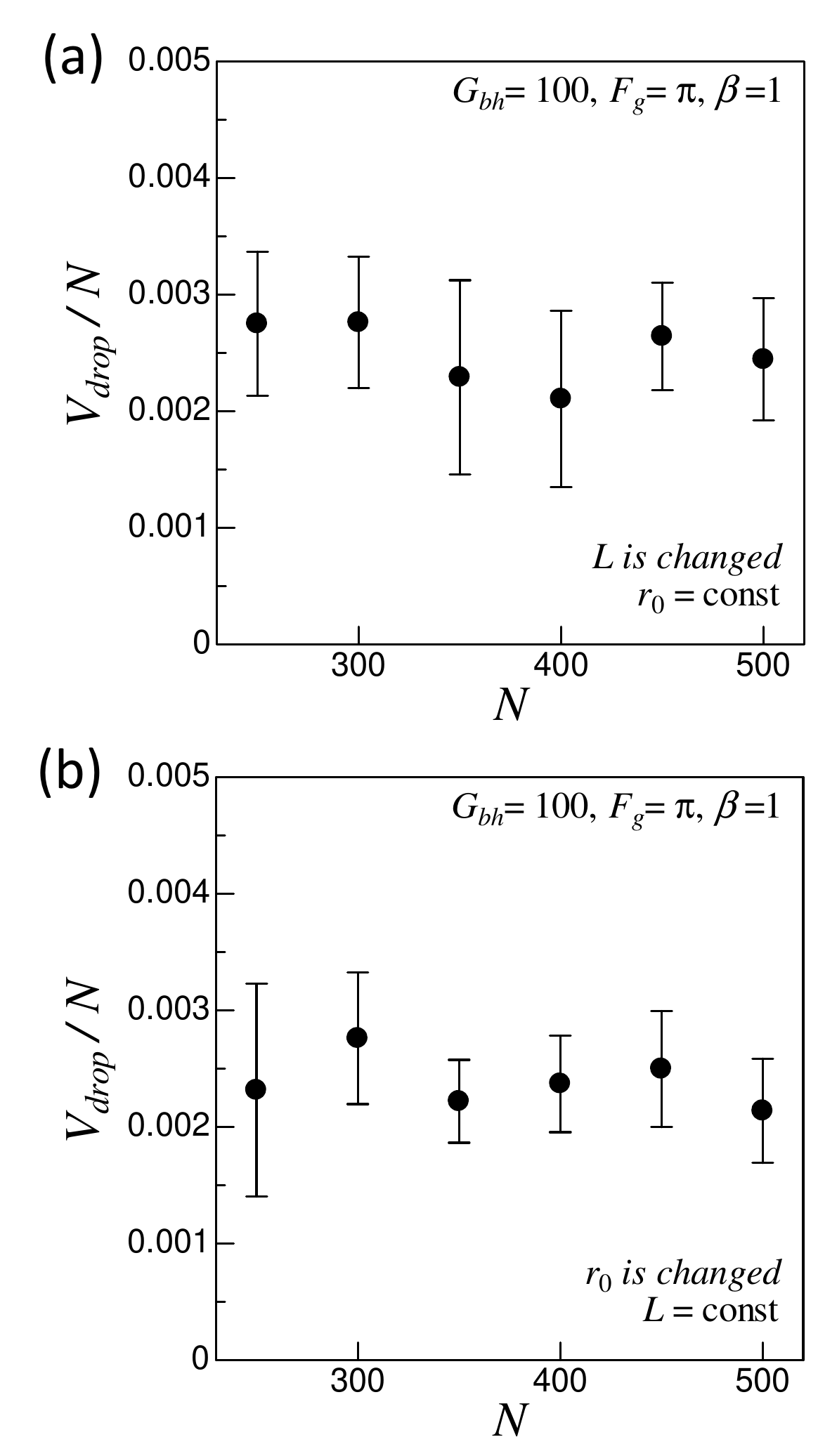}}
\caption{Effect of the number of particles $N$ on the growth rate $V_{drop}$ of the clustering instability of pullers ($\beta = 1, F_g = \pi, G_{bh} = 100$), where the velocity is normalized by $N$. (a) The vertical length $L$ is changed while the radius $r_0$ is kept constant. (b) The radius $r_0$ is changed while $L$ is kept constant.}
\label{fig5}
\end{center}
\end{figure}

The mechanism for this clustering instability can be understood by examining the hydrodynamic interactions between pullers, as schematically shown in Fig.~\ref{fig2}(a). When the pullers are aligned and directed in the vertical direction, they attract each other vertically but repel each other horizontally. Hence, the vertical jet shrinks to form a cluster. Once a cluster of pullers is formed, it is stable as shown in Fig.~\ref{fig1}.
The volume fraction of squirmers in the cluster is in the range of approximately 30-35\% (so an increase from its initial value), and the characteristic centre-centre separation between neighbour squirmers is approximately $2.5 \sim 2.6$. In contrast with past work in the dilute limit, the  stability of the cluster is revealed   here using a method   that  accurately accounts for near-field interactions.

To clarify the mechanism for such a dynamically stable active droplet, we plot the velocity vectors of the squirmers relative to their average velocity in Fig.~\ref{fig2}(b). Squirmers in the center of the droplet have downward velocities, whereas squirmers in the side of the droplet have upward velocities, leading to a dynamical vortex ring of squirmers. The sinking jet generates a background velocity field as shown in Fig.~\ref{fig2}(c). The vorticity field makes the squirmers face towards the center line of the jet. As a result, squirmers draw trajectories shown by the red arrows in Fig.~\ref{fig2}(c). The resulting cluster of pullers is then due to the interplay between the background flow and the squirming velocities.

\begin{figure*}[t]
\begin{center}
\centerline{\includegraphics[scale=0.48]{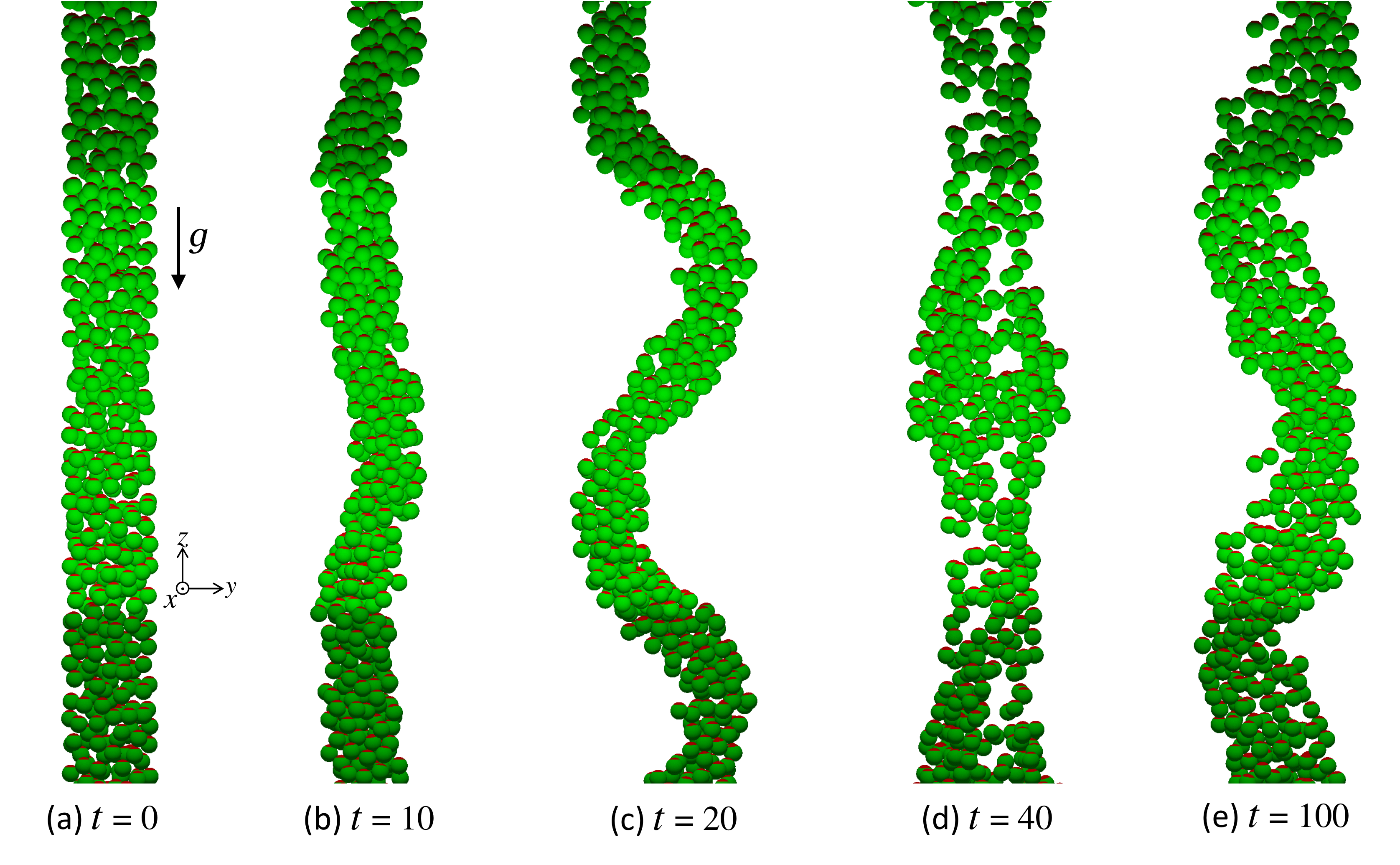}}
\caption{Waving instability displayed by pusher swimmers ($\beta = -1, F_g = 3\pi, G_{bh} = 100$ and $N=300$), where $t$ is non-dimensionalised by $a/U_0$; see also supplementary Movie 2~\cite{SM}. (a) Initially, 300 squirmers are placed in a column of radius $r_0$ with a vertically upward orientation, where gravity acts in the $-z$ direction. The anterior part of squirmer is colored in red. (b) Jet configuration at $t = 10$. (c) Jet configuration at $t = 20$, where the waving instability is observed. (d) Jet configuration at $t = 40$, where the waving instability is destroyed. (e) Jet configuration at $t = 100$, where the waving instability is again observed.}
\label{fig6}
\end{center}
\end{figure*}

In order to investigate the strength of the clustering instability, we calculate the growth rate of the perturbation $V_{drop}$. We divide the jet in the unit domain  into 20 vertical sections with equal intervals, and then count the number of squirmers in each section. Initially, the jet radius is constant along the $z$-axis, so the number of squirmers in each section is almost identical. When $t = 5$, we observe perturbations in the jet radius (see Fig.~\ref{fig1}(b)), and the number of squirmers in each section fluctuates as shown in Fig.~\ref{fig3}(a). In order to estimate the perturbation amplitude, a sinusoidal curve is fitted using least-squares. Then, we plot the time change of the amplitude as shown in Fig.~\ref{fig3}(b). We set a threshold amplitude of $A_c = 0.07N$, below which the amplitude increases steadily without large oscillations. The growth rate of the perturbation $V_{drop}$ is finally estimated as the slope of the time change of the amplitude, also using least-squares.

The impact of the parameters $G_{bh}, F_g$ and $\beta$ on the growth rate of the perturbation is shown in Fig.~\ref{fig4}. All data points indicate the average value of five independent simulations with different initial conditions, and the error bars measure the standard deviation in these five simulations. We see that the effect of $G_{bh}$  (the restoring torque on the cells) and $F_g$ (the net external force on the cells) on the growth rate is not significant. 
Although both gravity and bottom-heaviness are crucial in establishing the initial jet configuration, they do not appear to govern the instability observed numerically.

In contrast, the impact of the stresslet strength $\beta$ is significant (see Fig.~\ref{fig4}(c)), and the growth rate increases almost linearly with $\beta$ (the slope is fitted to be $\approx 0.562$). These results illustrate the predominant role of hydrodynamic interactions on the clustering instability, since  hydrodynamic interactions are strengthened as $\beta$ is increased.

The impact of the wavelength of the perturbation and jet radius on the growth rate of the perturbation is shown in Fig.~\ref{fig5}. In Fig.~\ref{fig5}(a), we vary the unit domain length $L$, with keeping initial jet radius constant, by changing $N$. In Fig.~\ref{fig5}(b), on the other hand, we change the initial jet radius $r_0$, while keeping $L$ constant, by changing $N$. Since the perturbation amplitude is evaluated by the number of squirmers, the growth rate tends to increase with $N$. To eliminate this effect, we normalize the growth rate by $N$ in Fig. \ref{fig5}. We see that the growth rate is not significantly affected in both cases. Therefore, the dominant control parameter for the clustering instability is $\beta$, i.e.~the hydrodynamic interactions between the squirmers.


\section{Waving instability induced by pushers}

Next, we investigate the waving instability induced by pushers. In Fig.~\ref{fig6} we  show the time change of the jet configuration for the parameters $\beta = -1, F_g = 3\pi, G_{bh} = 100$ and $N=300$ (see also supplementary Movie 2~\cite{SM}). We see that the jet configuration becomes wavy (sinuous mode) when $t = 10$. The waving instability grows with time, with a larger amplitude when $t = 20$. A too large amplitude destroys the wavy structure, and the jet returns to an almost straight configuration at $t = 40$. Then, the jet buckles again and displays the waving instability at $t = 100$. Such a waving instability,  not observed for Newtonian jets,  is a unique feature of  active jets;  it is reminiscent of the zigzag instability reported earlier for lines of pushers in the far field~\cite{Lauga2}.

\begin{figure}[t]
\begin{center}
\centerline{\includegraphics[scale=0.35]{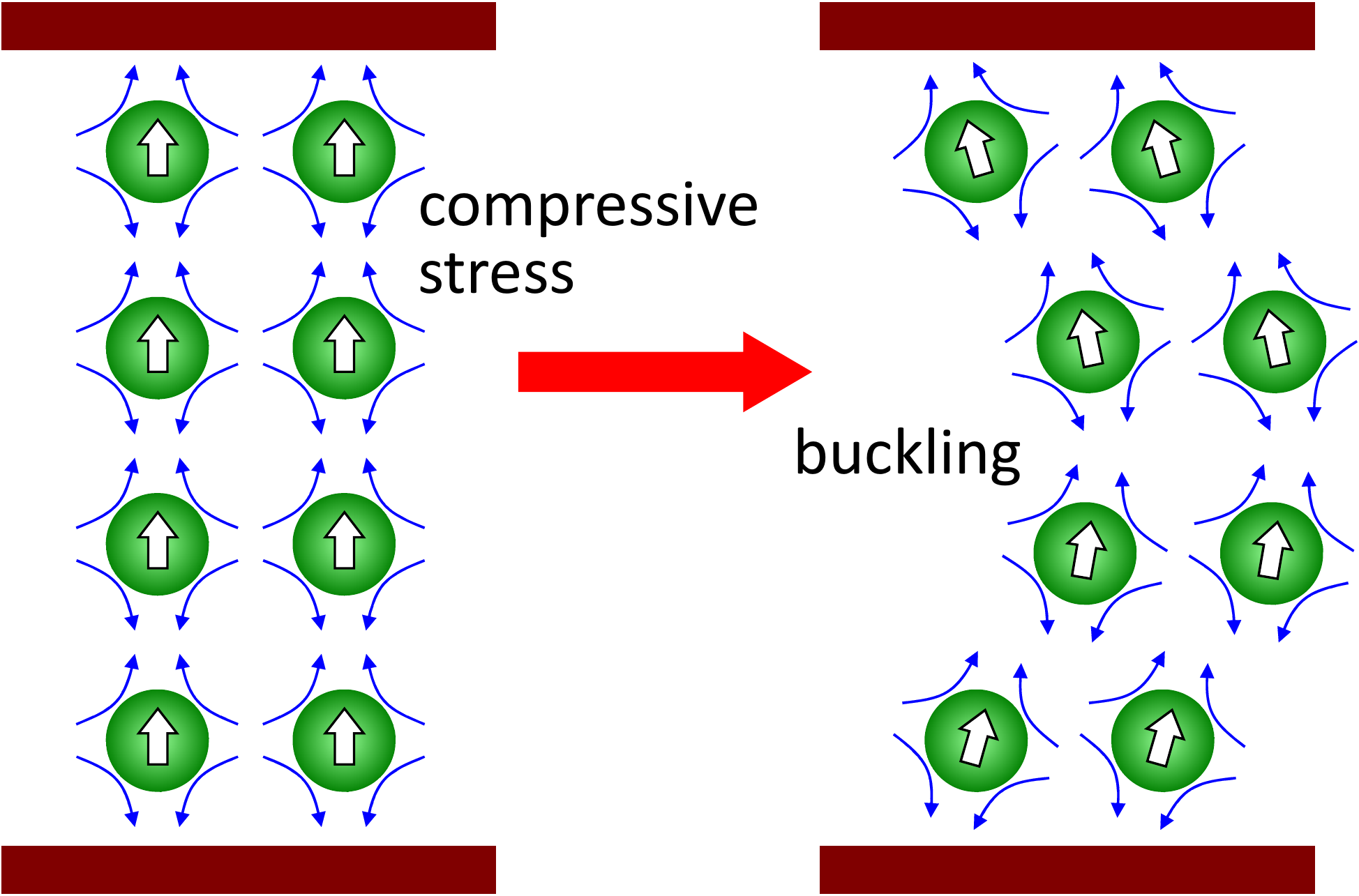}}
\caption{Schematic mechanism for the  waving instability of pushers. Squirmers vertically repel each other, and the compressive stresses induce the buckling of the assembly of swimmers.}
\label{fig7}
\end{center}
\end{figure}

The mechanism for this waving instability can be uncovered here again by examining the  hydrodynamic interactions between the pusher cells, as schematically shown in Fig.~\ref{fig7}. When pushers are aligned and directed  vertically upward, they repel each other vertically. Hence, a compressive stress is exerted along the jet. This compressive stress subsequently induces the buckling of the jet, leading to the waving deformation.

\begin{figure}
\begin{center}
\centerline{\includegraphics[scale=0.5]{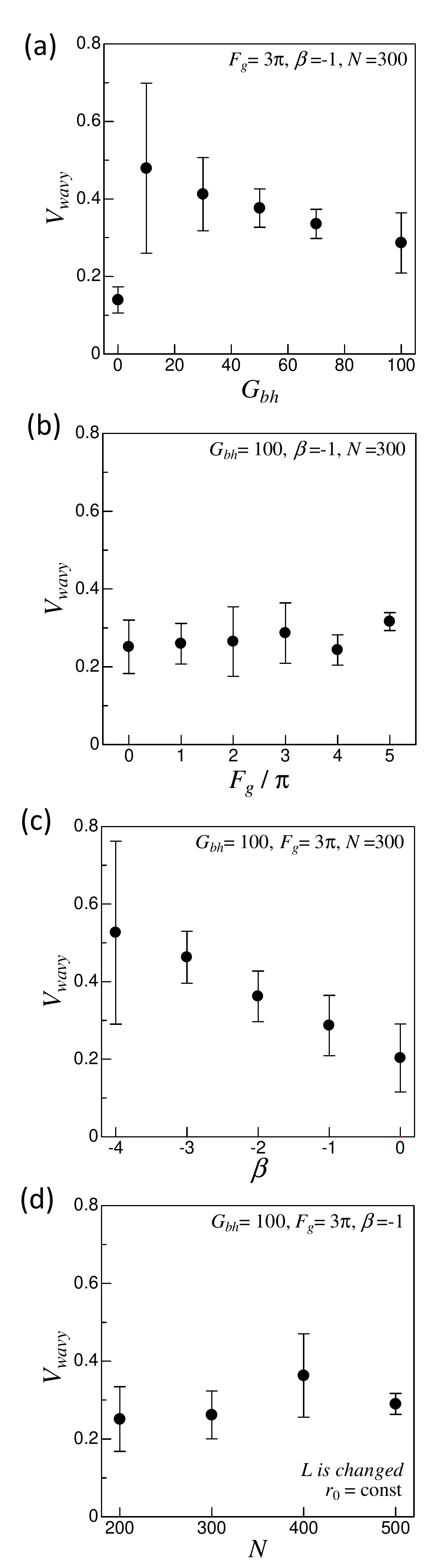}}
\caption{Effect of parameters on the growth rate $V_{wavy}$ of the  waving instability. Error bars indicate standard deviation of five different computations. (a) Effect of $G_{bh}$ ($\beta = -1, F_g = 3\pi, N = 300$). (b) Effect of $F_g$ ($\beta = -1, G_{bh} = 100, N = 300 $). (c) Effect of $\beta$ ($F_g = 3\pi, G_{bh} = 100, N = 300 $). (d) Effect of the number of particles $N$, where the vertical length $L$ is changed while the radius $r_0$ is kept constant.}
\label{fig8}
\end{center}
\end{figure}

In order to quantify the strength of the waving instability, we calculate next the growth rate of the perturbation, $V_{wavy}$. A jet in the unit domain is again divided into 20 vertical sections with equal intervals. The average position of squirmers in section $i$ is calculated as $\vec{R}_i = \sum_{j=1}^{N_i} \vec{r}_j / N_i$, where $N_i$ is the number of squirmers in section $i$ and $\vec{r}_j$ is the position of squirmer $j$. We also calculate the average position of all squirmers $\vec{R}_a$, and estimate the amplitude of the wavy configuration using the horizontal deviation defined as $\Delta R_i = |\vec{R}_i - (\vec{R}_i \cdot \vec{e}_z)\vec{e}_z - \vec{R}_a + (\vec{R}_a \cdot \vec{e}_z)\vec{e}_z|$, where $\vec{e}_z$ is the unit vector in $z$-direction. The time evolution of the amplitude is then plotted to estimate the growth rate of the waving instability. We set a threshold amplitude of $\Delta R_c = 4.2$, below which the amplitude increases steadily without large oscillations. The growth rate of the perturbation $V_{wavy}$ is finally estimated as the slope of the time change of amplitude using least squares.

The effects of parameters $G_{bh}, F_g$, $\beta$ and $L$ on the growth rate of the perturbation are shown in Fig.~\ref{fig8}. All data points indicate the average value of five independent simulations with different initial conditions, and the error bars measure the standard deviation of the five simulation cases. We see that the effects of $G_{bh}$ (restoring torque), $F_g$ (external force) and $L$ (domain size) on the growth rate are not significant. The growth rate slightly decreases in the range $10 \le G_{bh} \le 100$, because larger $G_{bh}$ prevents orientation change of squirmers to swim horizontally. As for pullers, both gravity and bottom-heaviness  establish the initial jet configuration but do not  govern the instability. In contrast, the effect of $\beta$ is  significant (see Fig.~\ref{fig8}(c)),  illustrating again that the waving instability is controlled by  hydrodynamic interactions.


\section{Discussion} 

In this study, we investigated new features of  the collective properties of active fluids. Specifically, we uncovered  instabilities arising in dense jets of active fluids, modelled as  suspension of bottom-heavy and heavy squirmers. Jets of pullers were shown to break up into active droplets (varicose mode), while in the pusher cases the jet buckles into a waving (sinuous) mode. Both situations were shown to originate from hydrodynamic interactions between the individual swimmers. 

Both cases can be contrasted to classical results for simple Newtonian fluids. A column of fluid with surface tension is known to be subject to the capillary (or Rayleigh-Plateau) instability and breaks into small droplets. The  wavelength for an inert fluid jet at which perturbations grow fastest is about nine times the jet radius~\cite{Rayleigh}. The instability we obtain for pullers is visually similar, but of course it does not follow the same physics. Here for the active fluid jet of pullers, the perturbation growth rate was almost independent from the system size $L$ (see Fig.~\ref{fig5}(a)). Moreover, in the case of the simple fluid column, the capillary instability results from a balance between surface tension and inertia in the fluid. For an active fluid jet, there is no inertia (we assume Stokes flows) and instead the dynamics results from 
 hydrodynamic interactions between active pullers. In the case of pushers, the waving instability results from compression generated in the assembly of swimmers, here also generated by hydrodynamic interactions and their stresslet fields.
The perturbation growth rate was again almost independent from the system size $L$ (see Fig.~\ref{fig8}(d)).

Although both gravity and bottom heaviness do play an essential role in realizing the downward jet, we found that they essentially play no role in the instabilities themselves, which are controlled by the stresslet flows. Why aren't the flows associated with the gravitational forces contributing?
It is straightforward to see that when point forces are  aligned in an infinite straight line, by reversibility of Stokes flows the corresponding Stokeslet flows are neither  attracting nor repelling. In contrast, when  point stresslets are  aligned in an infinite straight line,  they attract or repel each other (depending on their puller vs pusher nature). This is why the stresslet flows, although decaying faster in space than the Stokeslets due to gravity, control the  instabilities seen here.

Two past theoretical studies have reported work related to our computational results. 
A clustering instability arising in a line of aligned pullers moving along the same direction was analyzed theoretically~\cite{Lauga1}. This study found that hydrodynamic interactions between the swimmers lead to instabilities in density. The growth rate of the perturbation was derived to be proportional to the stresslet strength, i.e.~$\beta$, a theoretical result consistent with the present study (see Fig.~\ref{fig4}(c)). Jibuti \etal~\cite{Jibuti} also found numerically a clustering instability of pullers, mimicking phototactic microalgae in a Poiseuille flow. Although the problem setting in their work is different from the present study, the emerged phenomenon has some similarities. 
In the present study we demonstrated the stability of  squirmer  clusters, thanks to our Stokesian dynamics method that can accurately account for near-field interactions between squirmers.

More recently, a waving instability was reported for  a line of aligned pushers moving along the same direction~\cite{Lauga2}. That study  showed that the most unstable wavelength is equal to twice the inter-swimmer distance, and thus  the line of pushers gradually transformed into a zigzag line. Here again, the growth rate of the perturbation was found to be proportional to the dipole strength, i.e.~$|\beta|$, also consistent with the present study (Fig.~\ref{fig8}(c)). However, the wavelength found in the present study (jet) is clearly much larger than twice the inter-swimmer distance.  Moreover, we showed here  that the waving instability could appear repeatedly (see Fig.~\ref{fig6}), which has not been reported in former studies.
The mode that was most unstable under lined dilute conditions used in \cite{Lauga2} does not necessarily appear under random concentrated conditions as well, which could be the reason for the different results between the present study and \cite{Lauga2}.

Although the clustering and waving instabilities were observed here for pullers and pushers,  respectively, the results may take the  opposite form in different settings. The work of Ref.~\cite{Ishikawa7}  investigated the behaviour of bottom-heavy squirmers in a monolayer suspension. In their setting, puller squirmers tended to swim vertically upwards due to the bottom-heaviness, and they formed a horizontal band instead of a vertical jet. When bottom-heavy pullers are aligned horizontally, they repel each other horizontally. The resulting compressive stress induces the buckling of the band (this was actually observed, see Fig.~5 in~\cite{Ishikawa7}).
The repulsion in that case is similar mathematically  to the repulsion seen above in the case of pusher swimmers, and therefore the waving instability of a horizontal band of pullers in Ref.~\cite{Ishikawa7} is  analogous to that of the vertical jet of pushers studied in the current work.

\section*{Acknowledgments}

T.I. was supported by the Japan Society for the Promotion of Science Grant-in-Aid for Scientific Research (JSPS KAKENHI Grant No. 17KK0080, 21H04999 and 21H05308). T.I. and T.N.D. performed computations in Advanced Fluid Information Research Center, Tohoku University. This project has also received funding from the
European Research Council (ERC) under the European Union's Horizon 2020 research and innovation programme (grant agreement 682754 to E.L.)




\begin{thebibliography}{}

 \item[] $^*$ t.ishikawa@tohoku.ac.jp

 \bibitem[1]{Savart}
     Savart, F.,
      Memoire sur la constitution des veines liquides lancees par des orifces circulaires en mince paroi.
     Annal. Chim., 53, 337-398 (1833)

 \bibitem[2]{Rayleigh}
     Rayleigh, Lord, J. W. S.,
     On the Stability, or Instability, of certain Fluid Motions,
     Proc. London Math. Soc., 10, 4-13 (1879)

 \bibitem[3]{Eggers}
     Eggers, J.,
     Nonlinear dynamics and breakup of free-surface flows,
     Rev. Mod. Phys., 69, 865-929 (1997)

 \bibitem[4]{Lasheras}
     Lasheras, J. C., Hopfinger, E. J.,
     Liquid Jet Instability and Atomization in a Coaxial Gas Stream,
     Annu. Rev. Fluid Mech., 32, 275-308 (2000)

 \bibitem[5]{Eggers2}
     Eggers, J, Villermaux, E.,
     Physics of liquid jets,
     Rep. Prog. Phys., 71, 036601 (2008)

 \bibitem[6]{Polanco}
     Polanco, G., Holdo, A. E., Munday, G.,
     General review of flashing jet studies,
     J. Hazard. Mater., 173, 2-18 (2010)

 \bibitem[7]{Marchetti}
     Marchetti, M. C., et al.,
     Hydrodynamics of soft active matter,
     Rev. Mod. Phys., 85, 1143-1189 (2013)

 \bibitem[8]{Ramaswamy}
     Ramaswamy, S.,
     Active fluids,
     Nat. Rev. Phys., 1, 640-642 (2019)

 \bibitem[9]{Saintillan}
     Saintillan, D.,
     Rheology of Active Fluids,
     Annu. Rev. Fluid Mech., 50, 563-592 (2018)

 \bibitem[10]{Doostmohammadi}
     Doostmohammadi, A., Ignés-Mullol, J., Yeomans, J. M., Sagués, F.,
     Active nematics,
     Nat. Commun., 9, 3246 (2018)

 \bibitem[11]{Klotsa}
     Klotsa, D.,
     As above, so below, and also in between: mesoscale active matter in fluids,
     Soft Matter, 15, 8946 (2019)
     
\bibitem[12]{Tjhung}
     Tjhung, E., Nardini, C., Cates, M. E.,
     Phys. Rev. X, 8, 031080 (2018)

 \bibitem[13]{Bratanov}
     Bratanov, V., Jenko, F., Frey, E.,
     New class of turbulence in active fluids,
     Proc. Natl. Acad. Sci. U.S.A., 112, 15048-15053 (2015)

 \bibitem[14]{Ishikawa}
     Ishikawa, T.,
     Suspension biomechanics of swimming microbes,
     J. R. Soc. Interface, 6, 815-834 (2009)

 \bibitem[15]{Lopez}
     Lopez, H. M., et al.,
     Turning Bacteria Suspensions into Superfluids,
     Phys. Rev. Lett., 115, 028301 (2015)

 \bibitem[16]{Jibuti}
     Jibuti, L., et al.,
     Self-focusing and jet instability of a microswimmer suspension,
     Phys. Rev. E, 90, 063019 (2014)

 \bibitem[17]{Blake}
     Blake, J. R.,
     A spherical envelope approach to ciliary propulsion,
     J. Fluid Mech., 46, 199-208 (1971)

 \bibitem[18]{Pedley}
     Pedley, T. J.,
     Spherical squirmers: models for swimming micro-organisms,
     IMA J. Appl. Math., 81, 488-521 (2016)

 \bibitem[19]{Ishikawa2}
     Ishikawa, T., Pedley, T. J.,
     The rheology of a semi-dilute suspension of swimming model micro-organisms,
     J. Fluid Mech., 588, 399-435 (2007)

 \bibitem[20]{Ishikawa3}
     Ishikawa, T., Locsei, J. T., Pedley, T. J.,
     Development of coherent structures in concentrated suspensions of swimming model micro-organisms,
     J. Fluid Mech., 615, 401-431 (2008)

 \bibitem[21]{Ishikawa4}
     Ishikawa, T., Pedley, T. J.,
     Fluid particle diffusion in a semi-dilute suspension of model micro-organisms,
     Phys. Rev. E, 82, 021408 (2010)

 \bibitem[22]{Pedley1992}
     Pedley, T. J., Kessler, J. O.,
     Hydrodynamic Phenomena in Suspensions of Swimming Microorganisms,
     Annu. Rev. Fluid Mech., 24, 313-358 (1992)

 \bibitem[23]{Hill2005}
     Hill, N. A., Pedley, T. J.,
     Bioconvection,
     Fluid Dyn. Res., 37, 1-20 (2005)

 \bibitem[24]{Bees2020}
     Bees, M. A.,
     Advances in Bioconvection,
     Annu. Rev. Fluid Mech., 52, 449-476 (2020)

 \bibitem[25]{Kage2020}
     Kage, A., Omori, T., Kikuchi, K., Ishikawa, T.,
     The shape effect of flagella is more important than bottom-heaviness on passive gravitactic orientation in Chlamydomonas reinhardtii,
     J. Exp. Biol., 223, jeb205989 (2020)

 \bibitem[26]{Jekely}
     J\'{e}kely, G., Colombelli, J., Hausen, H. et al.,
     Mechanism of phototaxis in marine zooplankton,
     Nature, 456, 395-399 (2008)

 \bibitem[27]{Drescher}
     Drescher, K., Goldstein, R. E., Tuval, I.,
     Fidelity of adaptive phototaxis,
     Proc. Natl. Acad. Sci. U.S.A. 107, 11171-11176 (2010)

 \bibitem[28]{Maleprade}
     Maleprade, H., Moisy, F., Ishikawa, T., Goldstein, R. E.,
     Motility and phototaxis of Gonium, the simplest differentiated colonial alga,
     Phys. Rev. E, 107, 101, 022416 (2020)

 \bibitem[29]{Kaupp}
     Kaupp, U. B., Kashikar, N. D., Weyand, I.,
     Mechanisms of sperm chemotaxis,
     Annu. Rev. Physiol., 70, 93-117 (2008)

 \bibitem[30]{Stocker}
     Stocker, R., Seymour,J. R.,
     Ecology and physics of bacterial chemotaxis in the ocean,
     Microbiol. Mol. Biol. R., 76, 792-812 (2012)

 \bibitem[31]{Kessler}
     Kessler, J. O.,
     Hydrodynamic focusing of motile algal cells,
     Nature, 313, 218-220 (1985)

 \bibitem[32]{Lauga1}
     Lauga, E., Nadal, F.,
     Clustering instability of focused swimmers,
     Europhys. Lett., 116, 64004 (2017)

 \bibitem[33]{Lauga2}
     Lauga, E., Dang, T. N., Ishikawa, T.,
     Zigzag instability of biased pusher swimmers,
     EPL, 133, 44002 (2021)

 \bibitem[34]{Ishikawa5}
     Ishikawa, T., Simmonds, M. P. and Pedley, T. J.,
     Hydrodynamic interaction of two swimming model micro-organisms,
     J. Fluid Mech., 568, 119-160 (2006)

 \bibitem[35]{Ishikawa6}
     Ishikawa, T.,
     Vertical dispersion of model microorganisms in horizontal shear flow,
     J. Fluid Mech., 705, 98-119 (2012)

 \bibitem[36]{Banchio}
     Banchio, A. J., Brady, J. F.,
     Accelerated Stokesian dynamics: Brownian motion,
     J. Chem. Phys., 118, 10323-10332 (2003)

 \bibitem[37]{Beenakker}
     Beenakker, C. W. J.,
     Ewald sum of the Rotne–Prager tensor,
     J. Chem. Phys., 85, 1581-1582 (1986)

 \bibitem[38]{Ishikawa7}
     Ishikawa, T., Pedley, T. J.,
     Coherent Structures in Monolayers of Swimming Particles,
     Phys. Rev. Lett., 100, 088103 (2008)

 \bibitem[39]{SM}
See Supplementary Material Online

\end{thebibliography}
\end{document}